\documentclass[aps,eqsecnum,twocolumn,prd,showpacs]{revtex4}
\usepackage{graphicx}
\def\bold#1{\setbox0=\hbox{$#1$}%
     \kern-.025em\copy0\kern-\wd0
     \kern.05em\%\baselineskip=18ptemptcopy0\kern-\wd0
     \kern-.025em\raise.0433em\box0 }
\def\slash#1{\setbox0=\hbox{$#1$}#1\hskip-\wd0\dimen0=5pt\advance
         to\wd0{\hss\sl/\/\hss}}
\newcommand{\be}{\begin{equation}}
\newcommand{\ee}{\end{equation}}
\newcommand{\bea}{\begin{eqnarray}}
\newcommand{\eea}{\end{eqnarray}}
\newcommand{\nn}{\nonumber}

\newcommand{\ket}[1]{\left| #1 \right\rangle}

\begin{document}
\begin{titlepage}
\addtolength{\jot}{10pt}

\flushright{BARI-TH/09-611}

\title{\bf Rare $B_s$ decays to $\eta$ and $\eta^\prime$ final states}

\author{M. V.  Carlucci$^{a,b}$, P.  Colangelo$^a$, F.  De Fazio$^a$ }

\affiliation{ $^a$ Istituto Nazionale di Fisica Nucleare, Sezione di Bari, Italy\\
$^b$ Dipartimento di Fisica,  Universit$\grave{a}$  di Bari,
Italy}

\begin{abstract}
We study  exclusive $B_s$ decays to final states with  $\eta$ and $\eta^\prime$,  induced by the rare $b \to s \ell^+ \ell^-$ and $b \to s \nu \bar
\nu$ transitions. Differential decay rates and total branching fractions are predicted in the Standard Model,  adopting the
flavour scheme for the description of the $\eta$-$\eta^\prime$ mixing. We discuss the theoretical uncertainty related  to
the hadronic matrix elements. We also consider  these decay modes in a new Physics scenario with a single universal extra dimension, studying the dependence of
 branching ratios and decay distributions on the compactification scale $R^{-1}$ of the extra dimension.
 \end{abstract}

\vspace*{1cm} 

\maketitle
\end{titlepage}

\newpage
\section{Introduction}\label{sec:intro}
The study of $B_s$ properties and decays
plays an important role in the exploration of  the
Standard Model (SM) and  in the searches for new Physics phenomena, and a great experimental effort is   devoted  at present and  foreseen  in the near future
at the hadron colliders and at the B factories running at the $\Upsilon(5S)$ peak \cite{Anikeev:2001rk}.
In particular,  $B_s^0- {\bar B}_s^0$ oscillations  provide complementary
information with respect to $K^0$ and $B_d$ systems for the
analysis of CP violation.  CDF
and D0 Collaboration at the Fermilab TeVatron  have  recently carried out
measurements of  the  $B_s^0- {\bar B}_s^0$ mixing phase, by an angular analysis of  the  final state  $ J/\psi \, \phi $. The measured phase is   larger  than the SM
prediction   (although with a sizeable error) \cite{exp}.   If confirmed, this result would represent  an
evidence of Physics beyond SM.   For this reason, new measurements are foreseen,   involving also  the final states $J/\psi \, f_0(980)$,  $J/\psi \, \eta$ and $J/\psi \, \eta^\prime$;
the branching ratio of  $B_s \to J/\psi \, \eta$  has been recently measured by the Belle Collaboration  using the
$\eta \to \gamma \gamma$ and $\eta \to \pi^+ \pi^0 \pi^-$ modes to reconstruct  $\eta$ mesons
\cite{Drutskoy:2009ei}.

 $B_s$  is  of prime interest  also for  several rare
 decay modes, namely  those induced by the $b \to s$ transition,   that are potentially important  for detecting  new Physics effects. 
Here we focus  on the decays into $\eta$ and $\eta^\prime$  and a
pair of leptons,  either $\ell^+ \ell^-$ (with $\ell=e, \mu$ and $\tau$) or $\nu \bar \nu$. Our aim is 
 to give predictions for several observables  in the Standard Model,
discussing the theoretical uncertainties related to  the $\eta-\eta^\prime$ mixing and  to the hadronic matrix elements in the decay amplitudes.

A second purpose is to consider a specific new Physics scenario and study how the various
observables deviate from SM. The chosen framework is the Appelquist-Cheng-Dobrescu (ACD) model
 with a single universal extra dimension (UED) \cite{Appelquist:2000nn}. The model is  a minimal extension of
SM in $4+1$ dimensions,  with the extra dimension compactified to
the orbifold $S^1/Z_2$ and the fifth coordinate $y$ running from $0$ to
$2 \pi R$,   $y=0$ and $y=\pi R$  being fixed points of the orbifold. The fields  are allowed to propagate
 in all  $4+1$ dimensions, hence the model belongs to the class of  {\it universal} extra dimension scenarios. One of its  motivations
is the possibility of naturally providing candidates for the dark matter, an issue of fundamental importance.

In the ACD model  the SM particles correspond to  the zero modes of fields propagating in the
compactified extra dimension. In addition to the zero modes,    towers of
Kaluza-Klein (KK) excitations are predicted to exist, corresponding to the higher modes of the fields in the extra dimension;
such fields  are imposed to be even under a parity transformation in the fifth coordinate $P_5: y \to
-y$.  On the other hand,  fields  which are odd under $P_5$ propagate in the extra
dimension without zero modes, and correspond to particles  without  SM partners.

The masses of  KK particles  depend on  the radius $R$ of the compactified extra dimension,  the  new parameter with respect to SM
\footnote{The ACD Lagrangian may include, in addition to $5-d$ bulk terms, boundary terms introducing additional parameters in the theory. Such terms  are renormalized by
bulk interactions, but they are volume suppressed. A simplifying assumption is that they vanish at the cutoff scale, so that
the only new parameter with respect to SM is the radius $R$ of the compactified extra dimension.}.  For example,
  the  masses of the KK bosonic modes are given by:
 \be m_n^2=m_0^2+{n^2 \over R^2} \,\,\,\,\,\,\, n=1,2,\dots \ee
 $m_0$ being the mass of the zero mode, so that   for small values of $R$, i.e. at large compactification scales,  the KK particles decouple from the low energy regime.
 Another  property of the ACD model is  the conservation
of the KK parity $(-1)^j$,  $j$ being  the KK number.  KK parity
conservation implies  the absence of tree level contributions of
Kaluza Klein states to processes taking place at low energy, $\mu
\ll 1/R$,   forbidding the production of a single KK particle  off the interaction of standard particles.  This permits to use precise electroweak measurements  to provide a lower bound to the compactification scale:  
   ${1 / R} \ge 250-300$ GeV \cite{Appelquist:2002wb}.
 Moreover, this suggests  the possibility that the lightest  KK particles  are among  the  dark matter components, namely the $n=1$ 
Kaluza-Klein  excitations of the photon and  neutrinos  \cite{Cheng:2002iz,Hooper:2007qk}.

Since KK modes can affect the loop-induced processes,  Flavour Changing Neutral Current (FCNC) transitions are particularly
suitable for constraining this new Physics scenario,   and indeed  many observables are sensitive to the compactification radius in case, e.g., of processes with $B$ and $\Lambda_b$
  \cite{buras,noi,UEDvarie,UEDvarie1,UEDvarie2}.  Here we consider  rare $B_s$ decays into $\eta$ and $\eta^\prime$ mesons:
 $B_s \to
\eta^{(\prime)} \ell^+ \ell^-$ and $B_s \to \eta^{(\prime)} \nu
{\bar \nu}$, described by   the effective Hamiltonian reported in  Section \ref{hamiltonians}. We discuss in Section \ref{mixingFF} the role of the $\eta$-$\eta^\prime$
mixing and of the   $B_s \to
\eta^{(\prime)}$ form factors, and present  predictions 
for the various modes  in the following two Sections, with comments on  the   feasibility of measuring  properties of these processes.
Before concluding, we  discuss in the ACD model the $1/R$ dependence of  the $bs$ Unitarity Triangle, i.e. the condition among the  Cabibbo-Kobayashi-Maskawa (CKM)  elements involved in $B_s$
decays,  and in particular the possible  value of the CP violating phase $\beta_s$  in this model.

\section{Effective Hamiltonian for
$B_s \to \eta^{(\prime)} \ell^+ \ell^-$ and $B_s \to
\eta^{(\prime)} \nu {\bar \nu}$}\label{hamiltonians}

In the Standard Model, the effective $ \Delta B =-1$, $\Delta S =
1$ Hamiltonian describing  the  transition $b \to s \ell^+
\ell^-$ can be expressed in terms of a set of local operators:
\begin{equation}
H_{b \to s \ell^+ \ell^-}\,=-\,4\,{G_F \over \sqrt{2}} V_{tb}
V_{ts}^* \sum_{i=1}^{10} C_i(\mu) O_i(\mu) \label{hamil}
\end{equation}
\noindent where $G_F$ is the Fermi constant and $V_{ij}$ are
elements of the CKM mixing matrix
(terms proportional to $V_{ub} V_{us}^*$  are neglected since the ratio
$\displaystyle \left |{V_{ub}V_{us}^* \over V_{tb}V_{ts}^*}\right
|$ is ${\cal O}(10^{-2})$). The operators $O_i$ are written in
terms of quark and gluon fields:
\begin{eqnarray}
O_1&=&({\bar s}_{L \alpha} \gamma^\mu b_{L \alpha})
      ({\bar c}_{L \beta} \gamma_\mu c_{L \beta}) \nonumber \\
O_2&=&({\bar s}_{L \alpha} \gamma^\mu b_{L \beta})
      ({\bar c}_{L \beta} \gamma_\mu c_{L \alpha}) \nonumber \\
O_3&=&({\bar s}_{L \alpha} \gamma^\mu b_{L \alpha})
      [({\bar u}_{L \beta} \gamma_\mu u_{L \beta})+...+
      ({\bar b}_{L \beta} \gamma_\mu b_{L \beta})] \nonumber \\
O_4&=&({\bar s}_{L \alpha} \gamma^\mu b_{L \beta})
      [({\bar u}_{L \beta} \gamma_\mu u_{L \alpha})+...+
      ({\bar b}_{L \beta} \gamma_\mu b_{L \alpha})] \nonumber \\
O_5&=&({\bar s}_{L \alpha} \gamma^\mu b_{L \alpha})
      [({\bar u}_{R \beta} \gamma_\mu u_{R \beta})+...+
      ({\bar b}_{R \beta} \gamma_\mu b_{R \beta})] \nonumber \\
O_6&=&({\bar s}_{L \alpha} \gamma^\mu b_{L \beta})
      [({\bar u}_{R \beta} \gamma_\mu u_{R \alpha})+...+
      ({\bar b}_{R \beta} \gamma_\mu b_{R \alpha})] \nonumber \\
O_7&=&{e \over 16 \pi^2} m_b ({\bar s}_{L \alpha} \sigma^{\mu \nu}
     b_{R \alpha}) F_{\mu \nu} \nonumber \\
O_8&=&{g_s \over 16 \pi^2} m_b \Big[{\bar s}_{L \alpha}
\sigma^{\mu \nu}
      \Big({\lambda^a \over 2}\Big)_{\alpha \beta} b_{R \beta}\Big] \;
      G^a_{\mu \nu} \nonumber\\
O_9&=&{e^2 \over 16 \pi^2}  ({\bar s}_{L \alpha} \gamma^\mu
     b_{L \alpha}) \; {\bar \ell} \gamma_\mu \ell \nonumber \\
O_{10}&=&{e^2 \over 16 \pi^2}  ({\bar s}_{L \alpha} \gamma^\mu
     b_{L \alpha}) \; {\bar \ell} \gamma_\mu \gamma_5 \ell
\label{eff}
\end{eqnarray}
\noindent with $\alpha$, $\beta$  colour indices,
$\displaystyle b_{R,L}={1 \pm \gamma_5 \over 2}b$, and
$\displaystyle \sigma^{\mu \nu}={i \over
2}[\gamma^\mu,\gamma^\nu]$; $e$ and $g_s$ are the electromagnetic
and the strong coupling constant, respectively,  and $F_{\mu
\nu}$ and $G^a_{\mu \nu}$ in $O_7$ and $O_8$ denote the
electromagnetic and the gluonic field strength tensor. $O_1$ and
$O_2$ are current-current operators, $O_3,...,O_6$   QCD
penguin operators, $O_7$  and $O_8$  magnetic penguin operators, $O_9$ and
$O_{10}$  semileptonic electroweak penguin operators. The
Wilson coefficients  in (\ref{hamil}) have been computed
at NNLO in the Standard Model  \cite{nnlo}.  The operators $O_1$ and $O_2$  contribute to the the final state with a lepton pair
through a $\bar c c$ contribution that can give rise to charmonium resonances $J/\psi$, $\psi(2S)$, etc.
The resonant term can be controlled  and subtracted by appropriate
kinematical cuts around the  resonance masses. Since the Wilson coefficients $C_3-C_6$ are small,   the contribution of only the operators $O_7$, $O_9$ and $O_{10}$
can be kept for the description of the  $b \to s \ell^+  \ell^-$ transition, with a modification of the Wilson coefficient $C_7$ described below.

The SM effective Hamiltonian  for  $b \to s \nu \bar \nu$:
\begin{equation}
H_{b \to s\nu \bar \nu}= {G_F \over \sqrt{2}} {\alpha \over 2 \pi
\sin^2(\theta_W)} V_{tb} V_{ts}^* \eta_X X(x_t) \, O_L  = C_L
O_L\label{hamilnu}
\end{equation}
involves the operator
\begin{equation}
O_L = {\bar s}\gamma^\mu (1-\gamma_5) b {\bar \nu}\gamma_\mu
(1-\gamma_5) \nu \,\, .\label{opnu}
\end{equation}
 $\theta_W$  is the Weinberg angle; the  function $X(x_t)$
($x_t=\displaystyle{ m_t^2 \over M_W^2}$,  with $m_t$  the top
quark mass) has been computed in \cite{inami}  and
\cite{buchalla,urban}, while  the QCD factor $\eta_X$ is close  to one
\cite{buchalla,urban,Buchalla:1998ba}, so that we use $\eta_X=1$.

In the ACD model  no operators other than those in (\ref{hamil}), (\ref{eff}) 
and (\ref{hamilnu}) contribute to  $b \to s \ell^+ \ell^-$ and  $b \to s \nu \bar \nu$.
  The model  belongs to the class of Minimal Flavour Violating (MFV) models,
and the  effects beyond SM are only encoded in the Wilson coefficients of the effective Hamiltonian
  \cite{buras}. 
   KK excitations   modify  the  coefficients $C_i$ and $C_L$,  which  acquire a
depence on the  compactification scale $1/R$.  For large values
of $1/R$, due to decoupling of  massive KK states,
    the   coefficients $C_i$ and $C_L$ reproduce the  Standard Model values
and the SM phenomenology is  recovered.

The Wilson coefficients can be   expressed as  functions $F(x_t,1/R )$
generalizing  the SM analogues $F_0(x_t)$: \be
F(x_t,1/R)=F_0(x_t)+\sum_{n=1}^\infty F_n(x_t,x_n) \,, \label{fxt}
\ee
 with
$x_n=\displaystyle{ n^2 \over R^2 M_W^2}$.  Remarkably,   the sum over the KK
contributions in  (\ref{fxt}) is finite,   a
consequence of a generalized GIM mechanism \cite{buras};  the SM results are recovered for  $R
\to 0$,  since $F(x_t,1/R) \to F_0(x_t)$ in that limit.

For $1/R$  of the order of a few hundreds of GeV the coefficients differ from the Standard Model values, and the physical observables are predicted to
be different than in  SM.  For the exclusive decays, however, it is important to study if this  effect can be picked  up,  or it is obscured by other
uncertainties, in particular those affecting  the  $B_s$ and $\eta$,
$\eta^\prime$ matrix elements  of the operators in
(\ref{hamil}), (\ref{hamilnu}).

\section{$\eta$ -$\eta^\prime$ mixing and $B_s \to \eta^{(\prime)}$ Form Factors}
\label{mixingFF}

In the determination  of  the $B_s \to \eta^{(\prime)}$ matrix elements we need to account for the
$\eta-\eta^\prime$ mixing. This is usually described in
two  schemes:  the singlet-octet (SO) and  the quark
flavour (QF) basis; in each scheme  two mixing angles are involved \cite{Feldmann:1999uf}.

In the SO basis one defines the  $\eta$ and $\eta^\prime$-vacuum matrix elements  of axial-vector currents:
\be
\langle0 | J^a_{\mu 5} | P(p) \rangle = i f^a_P p_\mu  \,\,\,\, (a=1,8) \,\,\, \label{SO}
\ee
where $P=\eta, \eta^\prime$ and $J^1_{\mu 5}$, $J^8_{\mu 5}$ are  $SU(3)_F$ singlet and octet axial-vector quark currents.
The four hadronic parameters $f^a_P$  in (\ref{SO}) can be written in terms of two angles: $\theta_1$ and $\theta_8$,  and of two decay constants $f_1$ and $f_8$ of a pure singlet and octet flavour state.
In this scheme, non-vanishing values of the angles $\theta_{1,8}$, as well as the difference $f_8 \neq f_\pi$ are $SU(3)_F$ breaking effects, and the contribution of the axial $U(1)$ anomaly is encoded
 in the decay constant $f_1$.

On the other hand, in the QF basis one defines the axial-vector currents:
\be
J^q_{\mu 5}= {1 \over \sqrt{2} } \left( \bar u \gamma_\mu \gamma_5  u + \bar d \gamma_\mu \gamma_5 d\right) \,\,\,\,, \,\,\,\,
J^s_{\mu 5}=  \bar s \gamma_\mu \gamma_5  s
\ee
and  the matrix elements:
\be
\langle 0 | J^b_{\mu 5} | P(p) \rangle = i f^b_P p_\mu \,\,\,\, (b=q,s) \,\,\, . \label{QF}
\ee
Also in this case the four parameters  $f^b_P$ can be expressed in terms of two angles $\varphi_q$ and $\varphi_s$,
and of two  decay constants $f_q$ and $f_s$ of states without and with strangeness, respectively  \cite{Feldmann:1999uf}.
The difference between the mixing angles $\varphi_q-\varphi_s$ is due to OZI-violating effects and is found to be small  
($\varphi_q-\varphi_s  <  5^\circ$), so that it has been proposed that the approximation of describing the $\eta-\eta^\prime$ mixing in the QF basis and a single mixing angle
is  convenient \cite{Feldmann:1999uf}.
The simplification  $\varphi_q\simeq \varphi_s \simeq
\varphi$ is  supported by a QCD sum rule analysis
of the decays $\phi \to \eta\gamma$ and  $\phi \to \eta^\prime\gamma$
\cite{DeFazio:2000my}. 
Here we adopt the quark flavour basis and define:
 \bea \ket{\eta_q}&=&{1 \over \sqrt{2} } \left(
\ket{{\bar
u} u} +\ket{{\bar d} d}\right) \nonumber \\
\ket{\eta_s}&=& \ket{{\bar s} s} \,\, , \label{etaqs} \eea
so that the $\eta$-$\eta^\prime$ system can be described in terms of the  mixing angle  $\varphi=\varphi_q=\varphi_s$:
\bea 
\ket{\eta}&=&  \cos \, \varphi \ket{\eta_q}- {\sin} \, \varphi  \ket{\eta_s} \nonumber \\
\ket{\eta^\prime}&=& { \sin} \, \varphi  \ket{\eta_q}+{\cos} \, \varphi \ket{\eta_s}  \,\,\,\, . \label{mixing} 
\eea 

There are several ways to measure  $\varphi$, namely through the radiative
transitions involving a light vector meson $V$, such as $V \to
\eta^{(\prime)} \gamma$ or $\eta^\prime \to V \gamma$,  or  studying  the two-photon decays $\eta^{(\prime)} \to \gamma \gamma$
 \cite{escribano}.
A precise result has recently been obtained by the KLOE Collaboration which, measuring
 the ratio $\displaystyle{ \Gamma(\phi \to \eta^\prime  \gamma) \over \Gamma(\phi \to \eta \gamma)}$ and  assuming the flavour basis  with  a single mixing angle,  quotes:
$\varphi=\big( 41.5 \pm 0.3_{stat} \pm 0.7_{syst} \pm0.6_{th} \big )^\circ$  \cite{kloe}.  We use this  value  in our study.
In the same analysis, KLOE has also allowed for a glue content in
 $\eta^\prime$, modifying the second relation in (\ref{mixing}):
 \bea 
 \ket{\eta^\prime}&=& {\rm cos} \, \varphi_G \,{\rm sin} \, \varphi \ket{\eta_q}+ {\rm cos} \, \varphi_G \, \,{\rm cos} \,
\varphi \ket{\eta_s} \nn\\&+& {\rm sin}\, \varphi_G \ket{gluons}  \label{gluons} \eea
 where $ \varphi_G$  is the mixing angle for the glue contribution.
In this case, considering also the  results for  the decay widths  $\Gamma(\eta^\prime \to \gamma \gamma)$, $\Gamma(\eta^\prime \to \rho \gamma)$ and $\Gamma(\eta^\prime \to \omega \gamma)$,
together with $\Gamma(\pi^0 \to \gamma \gamma)$ and $\Gamma(\omega \to \pi^0 \gamma)$,
KLOE obtains: $\varphi=(39.7 \pm 0.7)^\circ$ and ${\rm cos}^2 \, \varphi_G=0.86 \pm 0.04$.
We  comment below on how this measurement affects our predictions.

The mixing parameters are useful to determine the other quantities needed for the description of $B_s \to \eta^{(\prime)} \ell^+ \ell^-  (\nu \bar \nu)$ transitions,   the
  $B_s$ and $\eta^{(\prime)}$ matrix elements of the operators
in (\ref{hamil}), (\ref{hamilnu}). As usual, such matrix elements  are parameterized in terms of form factors: \bea \hskip
-0.5cm&&<\eta^{(\prime)}(p^\prime)|{\bar s} \gamma_\mu b
|B_s(p)>=(p+p^\prime)_\mu F^{B_s \to \eta^{(\prime)}}_1(q^2) \nonumber
\\\hskip
-0.5cm&&+{M_{B_s}^2-M_{\eta^{(\prime)}}^2 \over q^2} q_\mu \left
(F^{B_s \to \eta^{(\prime)}}_0(q^2)-F^{B_s \to \eta^{(\prime)}}_1(q^2)\right ) \nn \\ \label{f0} \eea
\noindent ($q=p-p^\prime$, $F^{B_s \to \eta^{(\prime)}}_1(0)=F^{B_s \to \eta^{(\prime)}}_0(0)$) and
\bea \hskip
-0.5cm&&<\eta^{(\prime)}(p^\prime)|{\bar s}\; i\;  \sigma_{\mu
\nu} q^\nu b |B_s(p)>= \nonumber \\ \hskip
-0.5cm&&\Big[(p+p^\prime)_\mu q^2
-(M_{B_s}^2-M_{\eta^{(\prime)}}^2)q_\mu\Big] \; {F^{B_s \to \eta^{(\prime)}}_T(q^2) \over
M_{B_s}+M_{\eta^{(\prime)}}} \hskip 3 pt .\nn \\ \label{ft} \eea

No  QCD sum rule  or lattice QCD calculations of  $B_s \to \eta^{(\prime)}$ form factors (which probe  the $\bar s s$ content of  $\eta$ and $\eta^\prime$) are available, yet.
The quark flavour scheme   allows to relate
 the  $B_s \to \eta_s$   form factors to
the $B \to K$ ones, so that: $F^{B_s \to \eta}=-{\rm sin} \, \varphi  F^{B \to K}$ and 
 $F^{B_s \to \eta^\prime}= {\rm cos}\, \varphi  F^{B \to K}$
(for  a generic form factor $F$),  keeping the physical masses of $B_s$, $\eta$ and $\eta^\prime$.
It is possible to estimate the uncertainty connected  to flavour symmetry breaking, 
 considering  relations holding in the heavy quark  limit and in the chiral
limit for heavy-to-light transition form factors
\cite{Wise:1992hn}. Taking as an example the form factor $F_1$,  one can write for  $q^2 \to q^2_{max}$: 
 \be
F_1^{H \to P}= {{\hat F} g \sqrt{M_H} \over 2  f_P (E_P+\Delta_H)} \label{vmd}
\ee where $H$  and $P$ are  a heavy  and a light pseudoscalar meson, respectively, and $E_P$ is the energy of the $P$ meson in the $H$ rest frame.
 $\hat F$ is the heavy meson decay constant in
the heavy quark limit, which is, at leading order in the heavy
quark mass $m_Q$, independent of it (modulo logarithms). $g$
describes the effective coupling $H^*HP$, $H^*$ being the vector
meson with the same quark content as $H$; $\Delta_H=M_{H^*}-M_H$.  If the main flavour breaking  effects are in $M_H$,  $\Delta_H$,  $E_P$  and $f_P$,
the ratio
${F_1^{B_s \to \eta_s}/F_1^{B \to K}  }$,
using $f_K=159.8 \pm 1.4 \pm  0.44$ MeV \cite{PDG} and $f_s=(1.34 \pm 0.06) f_\pi$
\cite{Feldmann:1999uf},  differs from $1$  by about  $10\%$,   a reasonable estimate of
$SU(3)_F$ breaking in the form factors.

In the following,  we use two different sets of form factors, both obtained by  QCD sum rules
  \cite{Colangelo:2000dp}.
We refer to the first set, computed 
 using short-distance  QCD sum rules, as set A  \cite{Colangelo:1995jv},  and to the second set, computed
by  light-cone QCD sum rules,  as set B  \cite{Ball:2004ye}. The comparison
allows a discussion of  the  uncertainty related to  the hadronic matrix elements.
In both sets  the error of the form factors is given at  $q^2=0$  and then
extended to  the full range of momentum transfer: $4 m_\ell^2 \le q^2 \le (M_{B_s}-M_{\eta^{(\prime)}})^2$.

We also consider a third determination obtained applying light-cone QCD sum rules
within the Soft Collinear Effective Theory (SCET)
\cite{Bauer:2000yr}.  It is interesting to consider  this framework, since it allows to express
the form factors in terms of a single universal function $\xi^{B_a \to P}$ ($a$
light flavour index) through the
relations, holding in the heavy-quark limit and in the large energy limit of the light meson $P$:
\bea F_1^{B_a \to P}(q^2)&=& \xi^{B_a \to P}(n_+ \cdot p^\prime) \nonumber \\
F_0^{B_a \to P}(q^2) &=& {2 E^\prime \over M_{B_a}} \xi^{B_a \to 
P}(n_+ \cdot p^\prime)   \label{scet-rel}  \\
F_T^{B_a \to P}(q^2) &=& {M_{B_a} +M_P \over M_{B_a}} \xi^{B_a \to
P}(n_+ \cdot p^\prime) \,\,. \nn\eea
In  (\ref{scet-rel})  $n_+$ is a light-like four-vector with
components $n_+=(1,0,0,-1)$, and $p^\prime$ is the  momentum of the $P$ meson, with $E^\prime=p^{\prime 0}$.
 At first order in the light pseudoscalar meson  mass the  following relations hold:
 \bea
 (n_+ \cdot p^\prime)&=& 2 E^\prime -{M_P^2 \over 2 E^\prime} \nonumber \\
E^\prime &=& {M_{B_a}^2+M_P^2-q^2 \over 2 M_{B_a}}
\label{kin}\,\,.\eea
 We refer to this  third parameterization as to set C. Developing light-cone QCD sum rules in the framework of the SCET,
in Ref.\cite{DeFazio:2005dx} the universal function
$\xi^{B \to \pi}$ relative to $B \to \pi$ form factors was
derived. By the same method, a  determination  can be obtained  in the case of  kaon, and the resulting
function  $\xi^{B \to K}$  can be parameterized as:
\be \xi^{B \to K}(n_+ \cdot
p^\prime)=\xi^{B \to K}(m_B)\left[ -a+{b \over (n_+ \cdot
p^\prime)} +c \,(n_+ \cdot p^\prime) \right] \,\label{xiK} \ee
with 
\bea &&\xi^{B \to K}(m_B) = 0.335^{+0.078}_{-0.094} \nonumber \\
&& a= 2.418 \hskip 0.7cm b=13.765 \hskip 0.7cm c=0.154
\,\,.\hspace*{0.6cm} \label{parxiK} \eea 
The $B_s \to \eta^{(\prime)}$  form factors
belonging to set C can be written in terms of the universal function: 
\bea 
F^{B_s \to \eta}_1 &=& - \sin \varphi \, \xi^{B \to K} \nonumber \\
F_0^{B_s \to \eta} &=& - \sin \varphi \,{2 E^\prime \over M_{B_s}} \xi^{B \to K}\nn\\
F^{B_s \to \eta}_T &=& - \sin \varphi \, {M_{B_s} +M_\eta \over M_{B_s}}\, \xi^{B \to K}
\nonumber \\
F^{B_s \to \eta^\prime}_1 &=&  \cos \varphi \, \xi^{B \to K}  \\
F_0^{B_s \to \eta^\prime}&=&   \cos \varphi \,{2 E^\prime \over M_{B_s}} \xi^{B \to K}\nn\\
F^{B_s \to \eta^\prime}_T &=&   \cos \varphi \,{M_{B_s} +M_{\eta^\prime} \over M_{B_s}}\, \xi^{B \to K} \,\,, \nn\eea
keeping the physical masses of the particles involved in the transition.
Although such relations are  valid in the heavy quark limit at small values of the momentum transfer $q^2$, we extrapolate them in the full
kinematical range of $B_s \to \eta^{(\prime)}$ transitions.  

Relating $B_s \to \eta^{(\prime)}$  form factors to the $B \to K$ ones in the QF scheme implies the neglect of  flavour-singlet contributions, for example
the annihilation contribution involving the strange quark spectator  in $B_s$ and producing $\eta^{(\prime)}$ through two gluons. 
 This contribution  has been  considered as possibly responsible 
of the anomalous pattern of the
branching fractions of nonleptonic $B$ and $D$ to
 $\eta^\prime$ decays. In particular,  one could expect that such a kind of contributions
are more important for $\eta^\prime$
than  for  $\eta$,  due to the  coupling of
$\eta^\prime$ to gluons  driven by the $U(1)_A$ anomaly in QCD  (at the origin of the large $\eta^\prime$ mass compared to the masses of light pseudoscalar mesons).

However, the size of  this contribution is not firmly established. Denoting this gluonic contribution as
 $F_G$ and the quark contribution as $F_q$,   a generic form factor $F$  can be parameterized as:
\bea
F^{B_s \to \eta}= -F _q {f_q \over f_K} \sin \varphi +  F_G \big( {\sqrt{2 \over 3}}{f_q \over f_K} \cos\varphi -{1 \over \sqrt {3}}{f_s \over f_K} \sin \varphi  \big) \nn \\
F^{B_s \to \eta^\prime}= F _q {f_q \over f_K} \cos \varphi +  F_G \big( {\sqrt{2 \over 3}}{f_q \over f_K} \sin\varphi +{1 \over \sqrt {3}}{f_s \over f_K} \cos \varphi  \big) 
\nn \\ \label{gluoncontr}
\eea
in terms of the decay constants $f_q=(1.02 \pm 0.02) f_\pi$ \cite{Feldmann:1999uf}, $f_s$ and $f_K$; eqs.(\ref{gluoncontr}) show that the main effect is for the $\eta^\prime$ form factors.
In case of  $B$ meson transitions,  various  analyses  conclude that  the singlet  term  $F_G$ could be sizeable, thus  possibly contributing to the processes 
involving  $\eta^\prime$  \cite{neubert,kurimoto,ball}, however  the uncertainties are large.  To estimate  such contributions  the   matrix elements:
\be <0| A_{\mu}^a(z)A_{\nu}^b(0)| \eta^{(\prime)}(p)> \ee
are needed
 ($A_\mu^a $ is the gluon field, and a gauge factor has been dropped). These matrix elements can be written in terms of distribution amplitudes (DA)
 expanded in  Gegenbauer polynomials.  However, already the first coefficient (denoted by $B_2$) of the  expansion of the twist-two distribution amplitude  $\phi^G$ (assumed to
be the same for $\eta$ and $\eta^\prime$)  is  uncertain:  for example,  $B_2(1.4 \, {\rm GeV})=4.6 \pm 2.5$  results  from a  combined analysis of
the $\eta^\prime \eta \gamma$ form factor and of the inclusive decay  $Y(1S) \to \eta^\prime X$  \cite{ali}.  
In correspondence to this value a contribution of about 5$\%$ is estimated for  the form factor $F_1^{B \to \eta^\prime}(0)$ \cite{ball},  and only if  $B_2$ is varied in a wider range
($B_2=0 \pm 20$)  up to ${\cal O}(20\%)$  corrections are found  in case of  $\eta^\prime$,  while the corrections  remain small in the case of $\eta$ meson.
In the framework of SCET these contributions are found to be consistent with zero \cite{zupan}.  
A constraint of the singlet contribution using the widths of the semileptonic decays  $B^+ \to \eta \ell^+ \nu_\ell$ and
$B^+ \to \eta^\prime \ell^+ \nu_\ell$
 is hampered by the present  experimental errors. 

One can investigate the presence of gluonic contributions to the form factors governing  
 the semileptonic modes   $D_s \to \eta \ell^+ \nu_\ell$ and $D_s \to \eta^\prime \ell^+ \nu_\ell$. The 
measurements $BR(D_s \to \eta \ell^+ \nu_\ell)=(3.2\pm0.5)\times 10^{-2}$ and 
$BR(D_s \to \eta^\prime \ell^+ \nu_\ell)=(1.12\pm0.35)\times 10^{-2}$ \cite{PDG} can be combined in the ratio:
$R_{D_s}={BR(D_s \to \eta^\prime \ell^+ \nu_\ell)\over BR(D_s \to \eta \ell^+ \nu_\ell)}=0.35 \pm 0.12$.
This value for $R_{D_s}$  is reproduced by  central values of the mixing angle: $\varphi=41.5^\circ$ and of  the ratio of the
two pieces contributing to  $F_1^{D_s \to \eta^{(\prime)}}$  written in  (\ref{gluoncontr}): ${F_G(0) \over F_q(0)}=0$,  thus pointing to small values of the singlet terms.

A possibility for a phenomenological analysis of $B_s\to \eta, \eta^\prime$  transitions could consists in considering the effects of  an arbitrarily added  flavour singlet contribution 
to the form factors   \cite{geng}. We prefer to provide predictions which do not include additional terms in the form factors.  The predictions
are  accurate in case of $\eta$;   deviations  in the modes with  $\eta^\prime$ would first prompt investigations  focused on the  singlet  contributions.

\section{$B_s \to \eta \ell^+ \ell^-$ and $B_s \to \eta^\prime \ell^+ \ell^-$ }\label{ell}

The observables in these modes are the differential and total decay rates. The expression for  the differential width:

\begin{widetext}
\be {d\Gamma (B_s \to \eta^{(\prime)} \ell^+
\ell^-) \over d q^2} = {G_F^2|V_{tb}V_{ts}^*|^2
\alpha^2 \over 2^{9} \pi^5} {\lambda^{1/2}(M_{B_s}^2,M^2_{\eta^{(\prime)}},q^2)
\over M_{B_s}^3}\sqrt{1-{4 m_\ell^2 \over q^2}}{1 \over 3  q^2}p(q^2)
\label{spettro_etall}\ee 
($\lambda(x,y,z)=x^2+y^2+z^2-2 x y -2 x z -2 y z$) involves the function
\be p(q^2)=6m^2_\ell (M_{B_s}^2-M^2_{\eta^{(\prime)}})^2 |\tilde b(q^2)|^2+
\lambda(M_{B_s}^2,M^2_{\eta^{(\prime)}},q^2)\left[ (2 m_\ell^2+q^2)|\tilde c(q^2)|^2-(4
m_\ell^2-q^2)|\tilde a(q^2)|^2 \right] \label{spettro_etall_1} \ee
\end{widetext} 
 where $\tilde a$, $\tilde b$ and $\tilde c$ are
combinations of form factors and Wilson coefficients:
\bea \tilde a(q^2)&=& C_{10} \, F^{B_s \to \eta^{(\prime)}}_1(q^2) \nonumber \\
\tilde b(q^2) &=&  C_{10} \, F^{B_s \to \eta^{(\prime)}}_0(q^2)\label{abc} \\
\tilde c(q^2) &=& C_9 \, F^{B_s \to \eta^{(\prime)}}_1(q^2)\nn \\ &-& 2 (m_b+m_s) C_7^{eff} { F^{B_s \to \eta^{(\prime)}}_T(q^2) \over M_{B_s}+M_{\eta^{(\prime)}}}
\,\,\,. \hspace*{0.8cm} \nonumber \eea
In the expression of $\tilde c$ in (\ref{abc}) we introduced the coefficient $C_7^{eff}$, which is a renormalization
scheme independent combination of $C_7, C_8$ and $C_2$, given by a 
 formula that  can be found, e.g., in \cite{noi}.
Using: $m_b=4.8 \pm 0.2$ GeV,  $m_s=104^{+26}_{ -34}$ MeV, 
together with the values:
$M_{B^0_s}=5366.3\pm0.6$ MeV,
$\tau_{B^0_s}=(1.470^{+0.026}_{-0.027}) \times
10^{-12}$ s, $V_{tb}=0.999$ and $V_{ts}=(38.7\pm2.3) \times 10^{-3}$ quoted by
the Particle Data Group \cite{PDG}, we obtain the following
branching ratios in the SM, depending on  the set of
form factors:
\be
{ BR}(B_s \to \eta \ell^+ \ell^-) =
\left\{
\begin{array}{lll}
 (1.2 \pm 0.3) \times 10^{-7} \hskip 0.4cm {\rm set \, A} \\
 (2.6 \pm 0.7)  \times 10^{-7} \hskip 0.4cm {\rm set\, B} \\
 (3.4 \pm 1.8)  \times 10^{-7} \hskip 0.4cm {\rm set \, C} \hskip 2 pt \label{br-eta-ll}
\end{array}
\right.
\ee
\be
{BR}(B_s \to \eta^\prime \ell^+ \ell^-) =
\left\{
\begin{array}{lll}
 (1.1 \pm 0.3) \times 10^{-7} \hskip 0.4cm {\rm set \, A} \\
 (2.2 \pm 0.6)  \times 10^{-7} \hskip 0.4cm {\rm set\, B} \\
 (2.8 \pm 1.5)  \times 10^{-7} \hskip 0.4cm {\rm set \, C} \hskip 2 pt  \label{br-etap-ll}
\end{array}
\right.
\ee
for $\ell=e, \mu$,   and
\be
{BR}(B_s \to \eta \tau^+ \tau^-) =
\left\{
\begin{array}{lll}
 (3\pm 0.5) \times 10^{-8} \hskip 0.6cm {\rm set \, A} \\
 (8 \pm 1.5)  \times 10^{-8} \hskip 0.6cm {\rm set\, B} \\
 (10 \pm 5.5)  \times 10^{-8} \hskip 0.4cm {\rm set \, C} \hskip 2 pt \label{br-eta-tt}
\end{array}
\right.
\ee
\be
{BR}(B_s \to \eta^\prime \tau^+ \tau^-) =
\left\{
\begin{array}{lll}
 (1.55 \pm 0.3) \times 10^{-8} \hskip 0.4cm {\rm set \, A} \hspace*{0.8cm} \\
 (3.85 \pm 0.75)  \times 10^{-8} \hskip 0.2cm {\rm set\, B} \\
 (4.7 \pm 2.5)  \times 10^{-8} \hskip 0.6cm {\rm set \, C} \hskip 2 pt  \label{br-etap-tt}
\end{array}
\right.
\ee

\begin{figure*}[t]
 \includegraphics[width=0.24\textwidth]{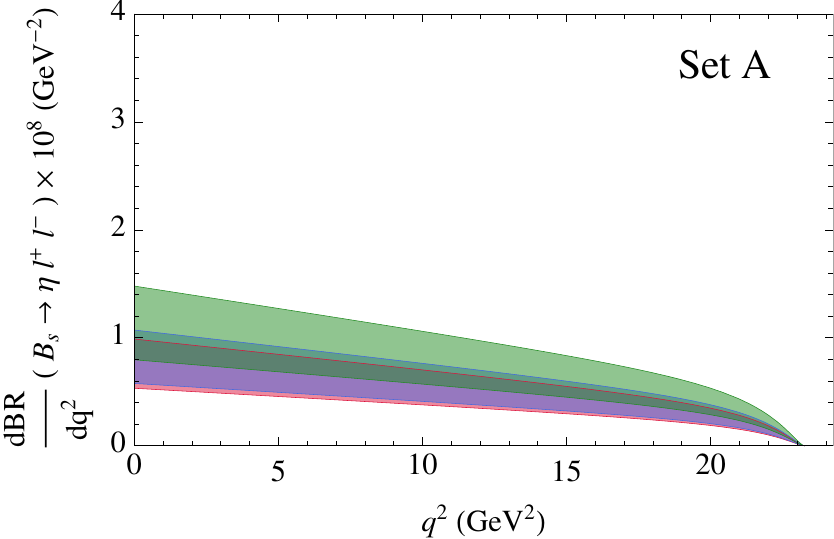} \hspace*{-0.10cm} 
 \includegraphics[width=0.22\textwidth]{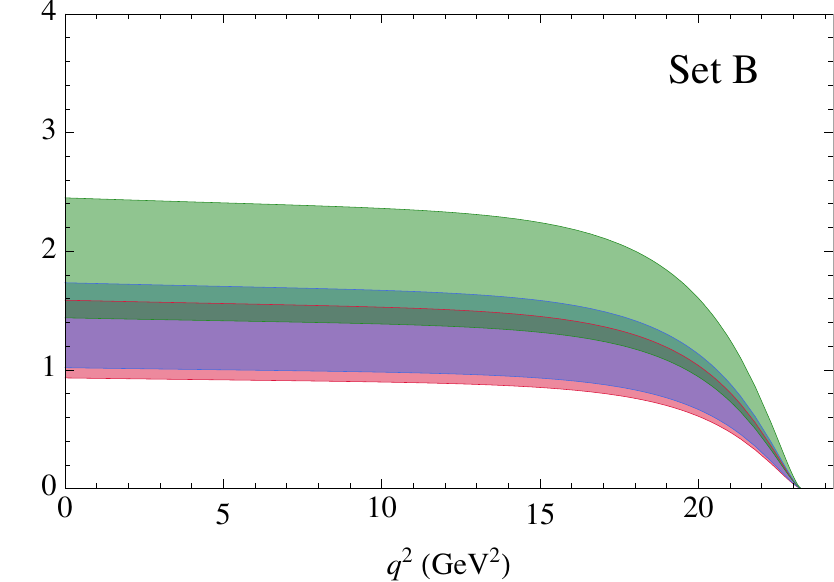}\hspace*{0.40cm} 
 \includegraphics[width=0.24\textwidth]{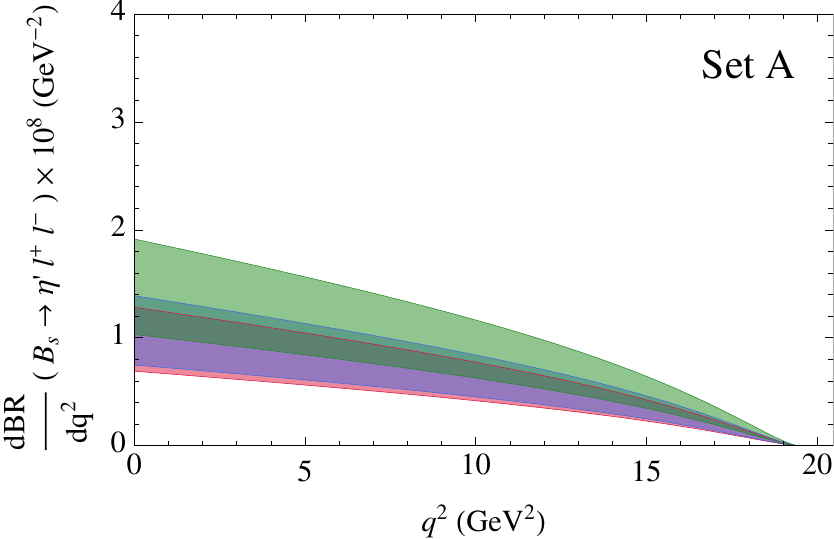} \hspace*{-0.2cm} 
 \includegraphics[width=0.22\textwidth]{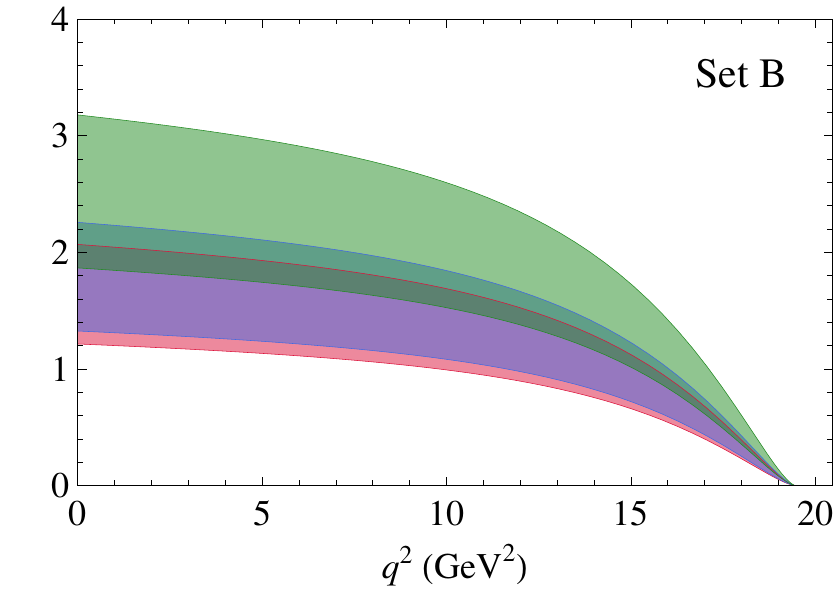}
\\ \vspace*{0.3cm}
\hspace*{0.1cm} \includegraphics[width=0.233\textwidth] {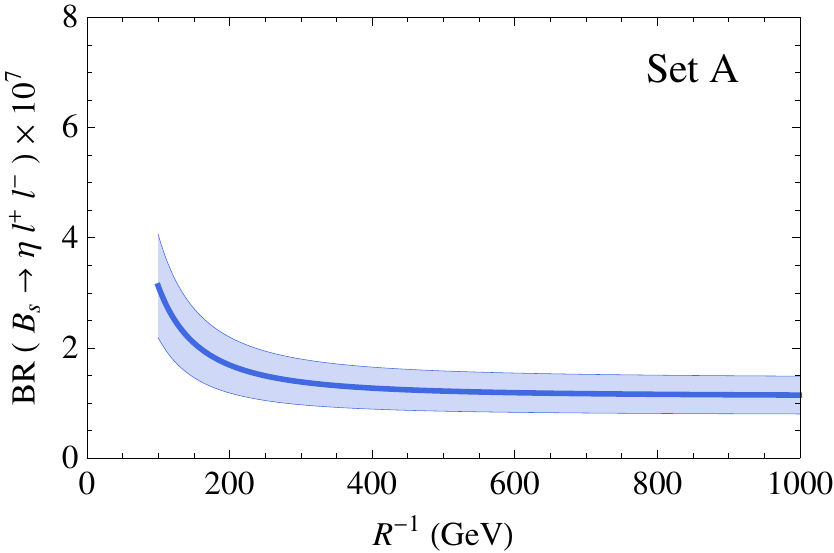}\hspace*{-0.1cm} 
 \includegraphics[width=0.225\textwidth] {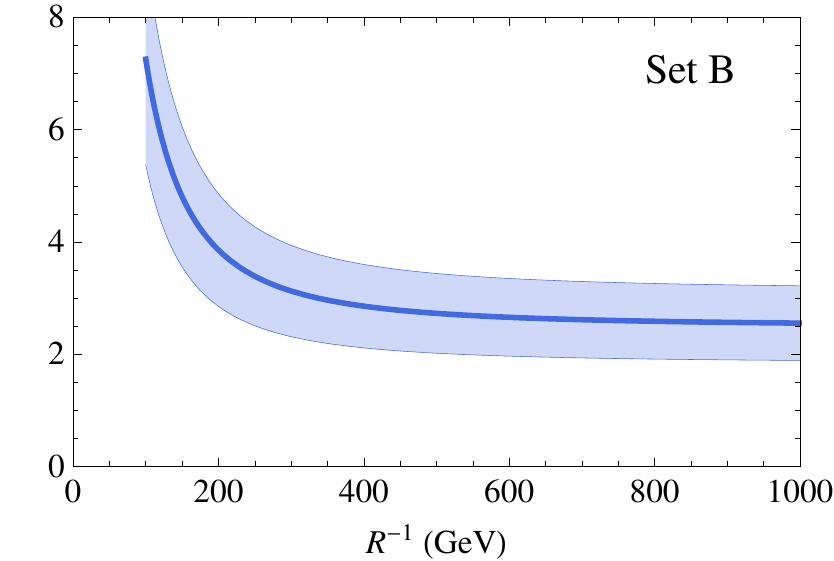}
\hspace*{0.4cm} \includegraphics[width=0.233\textwidth] {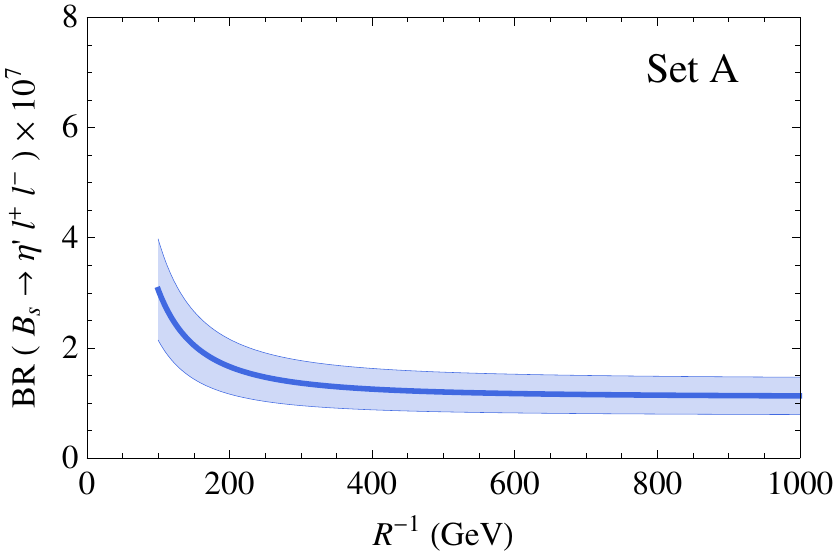}\hspace*{-0.2cm} 
 \includegraphics[width=0.225\textwidth] {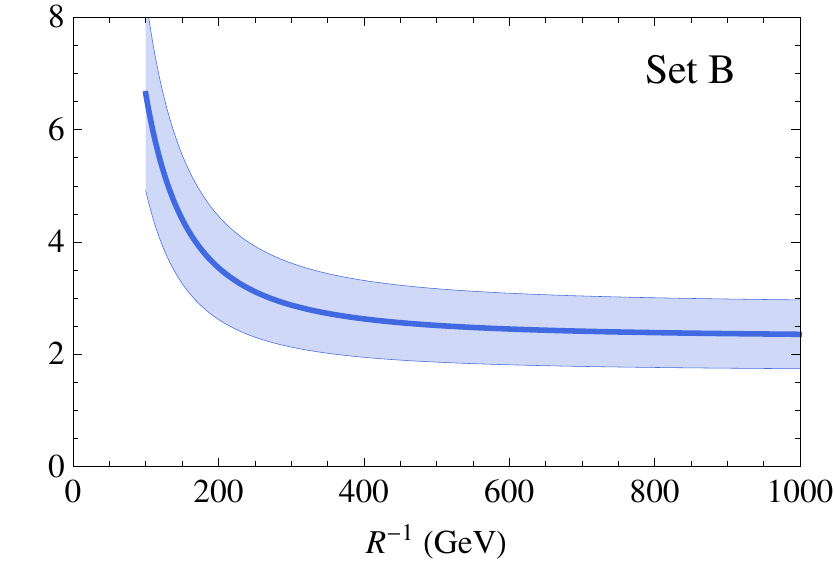}
\caption{
Differential branching fraction
${dBR(B_s \to \eta^{(\prime)} \ell^+ \ell^-) / dq^2}$  ($\ell=e, \mu$)  as a function of  the momentum transfer $q^2$ (upper plots),  and  branching fraction $BR(B_s \to \eta^{(\prime)} \ell^+ \ell^-)$
versus the compatification scale $\displaystyle{ R^{-1}}$ (lower plots)
obtained using set A    and B   of form factors.
In each  upper plot the lowest (red) region is obtained in  SM  ($1/R\to \infty$),
 the intermediate (blue) region for  the compactification scale $1/R=500$ GeV and the uppermost (green) region  for $1/R=200$ GeV. } \vspace*{1.0cm}
\label{spettroetall}
\end{figure*}

The results obtained  using the form
factor $\xi^{B \to K}$ in the SCET framework are affected by the
largest uncertainty and encompass the predictions based on  the
set A and set B. We  consider these results  as the most
conservative determinations of the decay rates,  as well as for the other observables.

The results from  set A and B are compatible within the errors, with lower central values  
for set A. The uncertainties,   estimated at the level of 30\% in the rates for each set of form factor, become  higher
when the central values are compared: the difference  provides us with hints on  the level of improvement in the determination of the form factors needed at present.
Such an improvement also concernes other determinations of the form factors  \cite{Skands:2000ru}. 

As a further remark, we notice that  the KLOE fit of the $\eta-\eta^\prime$ mixing eq.(\ref{gluons})
corresponds to using an effective mixing angle $\varphi^{eff}=(44.5\pm1.5)^\circ$ in the expressions of the $B_s \to \eta^\prime$ form factors,  and to
a $10\%$   decrease  of  the corresponding  decay rates.

 It is worth  comparing  the predictions  (\ref{br-eta-ll})-(\ref{br-etap-ll}) with the  results quoted by the
Heavy Flavour Averaging Group (HFAG) for the analogous  $B$ decay modes  \cite{HFAG}:
${BR}(B^0 \to K^0 \ell^+ \ell^-) =\big( 3.2^{+0.8}_{-0.6} \big) \times 10^{-7}$ and
${BR}(B^+ \to K^+ \ell^+ \ell^-) =(4.9\pm 0. 5) \times 10^{-7}$
for $\ell=e, \mu$.  On the other hand, in the approach based on the QF mixing scheme,  a relation connecting the decay width of $B_s \to \eta \ell^+ \ell^-$  to
that of $B_s \to K^0  \ell^+ \ell^-$ induced by the transition $b \to d$:
\be
{BR(B_s \to K^0 \ell^+  \ell^-) \over BR(B_s \to \eta \ell^+  \ell^-) } =  {1 \over (\sin \varphi)^2}  \left| {V_{td} \over V_{ts} }\right|^2 (1+\Delta_{SU(3)}+\Delta^\prime)\,\,\, ,
\ee
involves   an $SU(3)$ correction term $\Delta_{SU(3)}$ (presumably small)  and  a  term $\Delta^\prime$  coming from the additional contribution  proportional to $V_{ub} V^*_{ud}$,
the correspondent of which has been neglected in the effective Hamiltonian (\ref{hamil}) in case of $b\to s$ modes.  Such a kind of relations is
 of great interest,  both from the theoretical and the experimental viewpoint, and deserves a dedicated analysis.

In the ACD model  the Wilson coefficients $C_7^{eff}$, $C_9$ and $C_{10}$  depend on the compactification scale $1/R$, hence  decay rates and distributions vary
with  this parameter. 
$C_7^{eff}$ depends on $1/R$ through two functions
$D^\prime(x_t,{1 \over R})$ and $E^\prime(x_t,{1 \over R})$ 
representing  the contribution of $\gamma$-penguins and chromomagnetic
penguins, respectively. The contribution of $Z^0$-penguins enters
in the functions $Y(x_t,{1 \over R})$, $Z(x_t,{1 \over R})$ and
$X(x_t,{1 \over R})$:  $Y$ and $Z$ determine the coefficient $C_9$,
while $C_{10}$ depends only on $Y$. As for the function $X$, it enters in the
ACD expression of the coefficient $C_L$ relevant for the modes with neutrinos
discussed in the next Section. 

These functions obey the
general representation (\ref{fxt}),  and can be  found, e.g.,  in
\cite{buras,noi}. For values of the compactification scale $1/R$ of a few hundreds of GeV  the coefficient $C_7^{eff}$ is
suppressed in the ACD model with respect to the SM value, while
$C_{10}$ is enhanced and $C_9$ turns out to be almost unaffected. For
example, at ${1 \over R}=300$ GeV one has:
$\displaystyle{C_7^{eff,ACD} \over C_7^{eff,SM}} \simeq
0.82$,$\displaystyle{C_9^{ACD} \over C_9^{SM}} \simeq 1.01$,
$\displaystyle{C_{10}^{ACD} \over C_{10}^{SM}} \simeq 1.16$.  As a consequence,
for   the decay into $\eta$ and  $m_\ell=0$  we find for the function $p(q^2)$ in (\ref{spettro_etall_1})
at $q^2=q^2_{max}$ and $1/R=300$ GeV:
$\displaystyle{p(q^2_{max})|_{1/R=300 \, {\rm GeV}} \over
p(q^2_{max})|_{SM}} = 1.23-1.42$, depending on the set of form factors.  

The dependence of the rates on $1/R$ is obscured by the uncertainty of the hadronic form factors, as  shown in 
Fig. \ref{spettroetall},   at values of this scale  excluded by the  analyses of the electroweak constraints:  $1/R \ge 250$ GeV.  Sensitivity of  the decay rate to $1/R$
 could be achieved by a reduction in the hadronic uncertainty of about a factor of two:  in this case, the measurement of the rates of $B_s \to \eta^{(\prime)} \ell^+ \ell^-$ 
would allow to exclude the $1/R$ region below  $350-400$  GeV,  keeping  the same  uncertainties quoted for  the other input parameters.
Concerning the differential distributions ${d \Gamma \over d q^2}(B_s
\to \eta \ell^+ \ell^-)$ and ${d \Gamma \over d q^2}(B_s \to
\eta^\prime \ell^+ \ell^-)$  also depicted in  Fig.\ref{spettroetall},  the enhancement at small values of dilepton invariant mass obtained for decreasing  compactification scales $1/R$ 
could  be observed  after the same improvement of  the  accuracy of the form factors.

A possibility to reduce the effects of the hadronic uncertainties consists in comparing the decay distributions of  $B_s \to \eta \ell^+ \ell^-$  and $D_s \to \eta \ell^+ \nu_\ell$  ($\ell=e, \mu$)
  at large values of momentum transfer $q^2 \simeq q^2_{max}$,  i.e. close to the end-point. In this range the ratio:
\be
R(y)={{d\Gamma(B_s \to \eta \ell^+ \ell^-) / dy}\over  {d\Gamma(D_s \to \eta \ell^+ \nu_\ell) / dy}}
\label{ry}
\ee
obtained evaluating the distributions at the same value of 
$y={E \over M_\eta}$
can be written,  up to corrections  ${\cal O} ( {1 \over m_c} -{1 \over m_b} )$,  as:
\bea
R(y)&=&\left| {V_{tb}  V^*_{ts} \over V_{cs} }\right|^2 {\alpha^2 \over 8 \pi^2} \left( {M_{B_s} \over M_{D_s}Ê}  \right)^2 \nn \\
&&\left( |C_{10}|^2 + \left|C_9-{ 2 (m_b +m_s) \over M_{B_s}+M_\eta}  c_7^{eff} {F_T^{B_s \to \eta}(y)  \over
F_1^{B_s \to \eta}(y) }\right|^2 \right) \nn \\
\eea
thus providing an access to the Wilson coefficients $C_{7,9}$ and $C_{10}$, since the ratio of $F_T^{B_s \to \eta}(y)  \over
F_1^{B_s \to \eta}(y)$ is practically independent of $y$ and of  the set of form factors. 
To further reduce the impact of the hadronic corrections, the double
ratio could be considered:
\be
\tilde R(y)=
\left[{{d\Gamma(B_s \to \eta \ell^+ \ell^-) / dy} \over  {d\Gamma(D_s \to \eta \ell^+ \nu_\ell) / dy}}\right] /\left[
{{d\Gamma(B_s \to K \ell^+ \nu) / dy} \over  {d\Gamma(D_s \to K \ell^+ \nu_\ell) / dy}}\right]\\
\label{rtilde}
\ee
which is given by
\bea
\tilde R(y)&=&\left| {V_{tb}  V^*_{ts} \over V_{cs} }\right|^2  \left| {V_{cd} \over V_{ub} }\right|^2 {\alpha^2 \over 8 \pi^2} \left( {M_{B_s} \over M_{D_s}Ê}  \right)^2 \nn \\
&&\left( |C_{10}|^2 + \left|C_9-{ 2 (m_b +m_s) \over M_{B_s}+M_\eta}  c_7^{eff} {F_T^{B_s \to \eta}(y)  \over
F_1^{B_s \to \eta}(y) }\right|^2 \right) \nn \\
\eea
up to corrections  ${\cal O} (m_s \left({1 \over m_c} -{1 \over m_b} \right))$ \cite{grinstein}. However, the double ratio involves  many CKM suppressed transitions, and 
its  measurement is challenging.

\section{ $B_s \to \eta \nu \bar \nu$ and $B_s \to \eta^\prime \nu \bar \nu$  }\label{nu}

The exclusive decays with two neutrinos in the final state, which   from the theoretical side are among the cleanest FCNC processes,
require only one hadronic form factor. The observables are the decay distributions and the total decay widths. For the former,
 it is convenient to use the variable $\displaystyle x={E_{miss} \over M_{B_s}}$,
with  $E_{miss}$ the energy of the
neutrino pair in the $B_s$ rest frame  (missing energy), with $\displaystyle {0.5\leq x\leq 1-{M_{\eta^{(\prime)}}\over M_{B_s}}}$.
 In terms of the variable $x$ the differential decay rate $\displaystyle{d \Gamma \over dx}$ reads:
\bea
{d \Gamma(B_s \to \eta^{(\prime)} \nu \bar \nu) \over
dx} \hspace*{5cm} \nn \\
= 3\;{ |C_L |^2 \,|F^{B_s \to \eta^{(\prime)}}_1(q^2)|^2 \over 48 \pi^3 M_{B_s}}
\sqrt{\lambda^3(q^2, M_{B_s}^2, M_{\eta^{(\prime)}}^2)}\;,\hspace*{0.5cm}
\eea
with the invariant mass $q^2$ of the neutrino pair expressed in terms of $x$:   $q^2=M_{B_s}^2 (2 x-
1)+M_{\eta^{(\prime)}}^2$.  $C_L$ is the Wilson coefficient  in
(\ref{hamilnu})  and the factor of $3$ corresponds to  the sum
over the three neutrino flavours.

\begin{figure*}[ht]
\hspace*{-0.1cm}\includegraphics[width=0.24\textwidth] {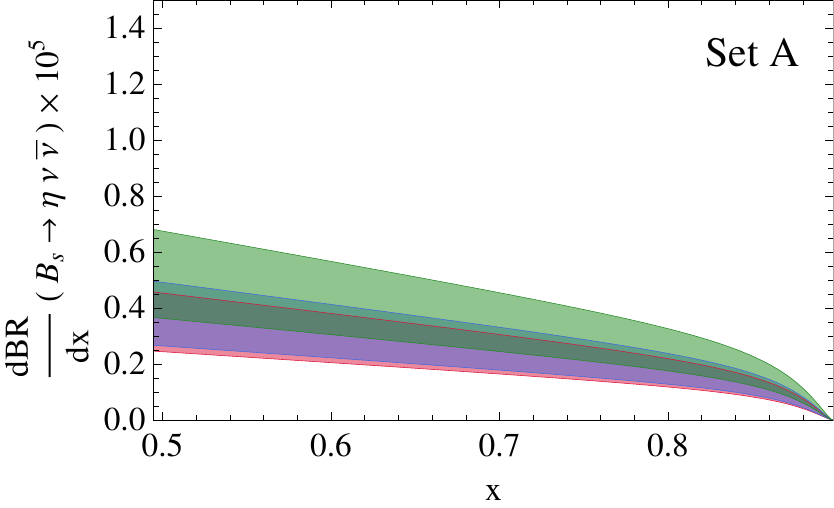}\hspace*{-0.1cm}
 \includegraphics[width=0.22\textwidth] {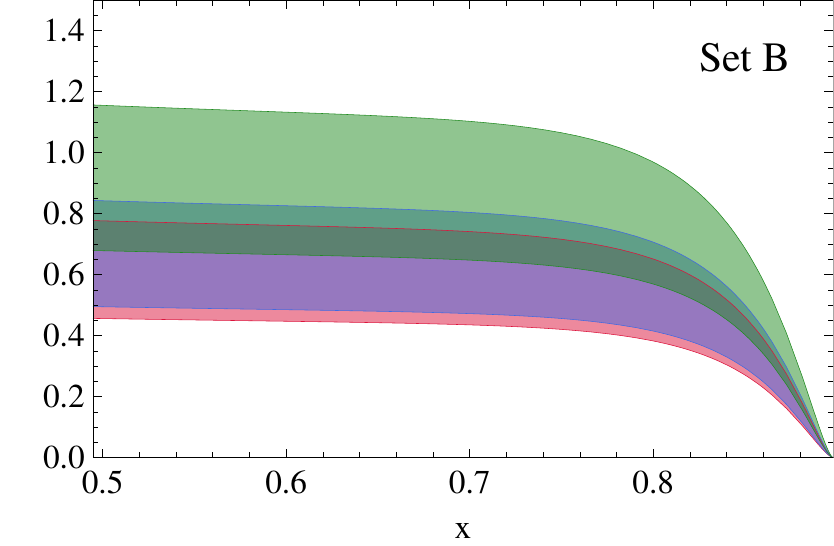}\hspace*{0.4cm}
 \includegraphics[width=0.24\textwidth] {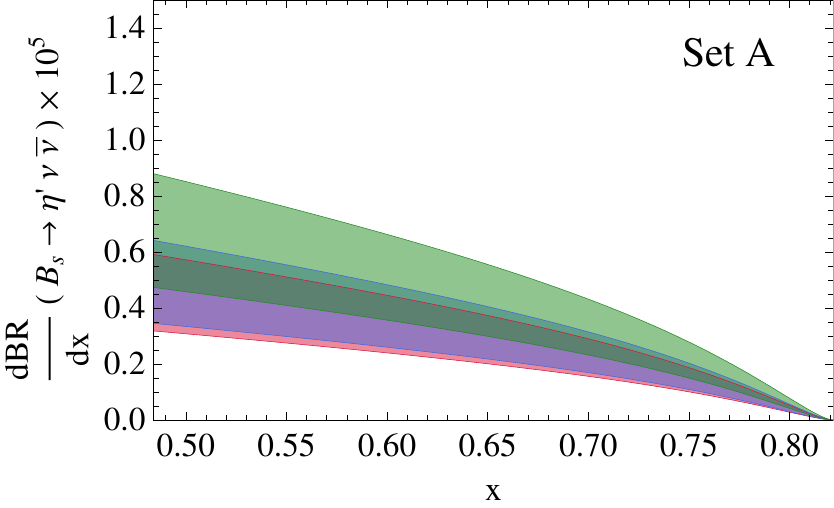}\hspace*{-0.1cm}
 \includegraphics[width=0.22\textwidth] {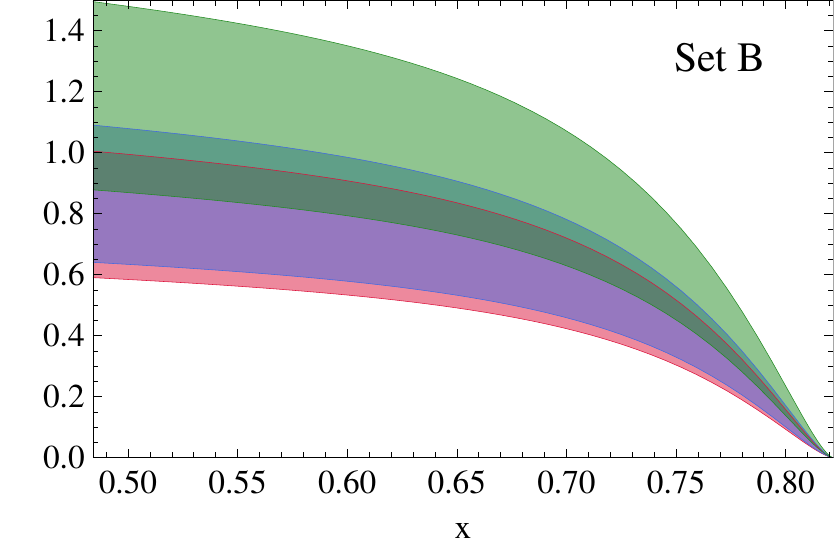}
 \\ \vspace*{0.3cm}
\hspace*{0.3cm} \includegraphics[width=0.233\textwidth] {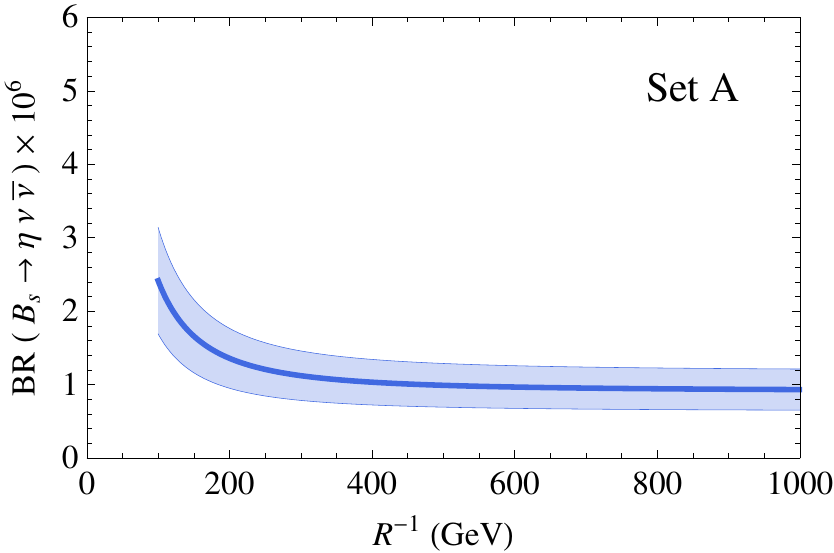}\hspace*{-0.1cm}
  \includegraphics[width=0.225\textwidth] {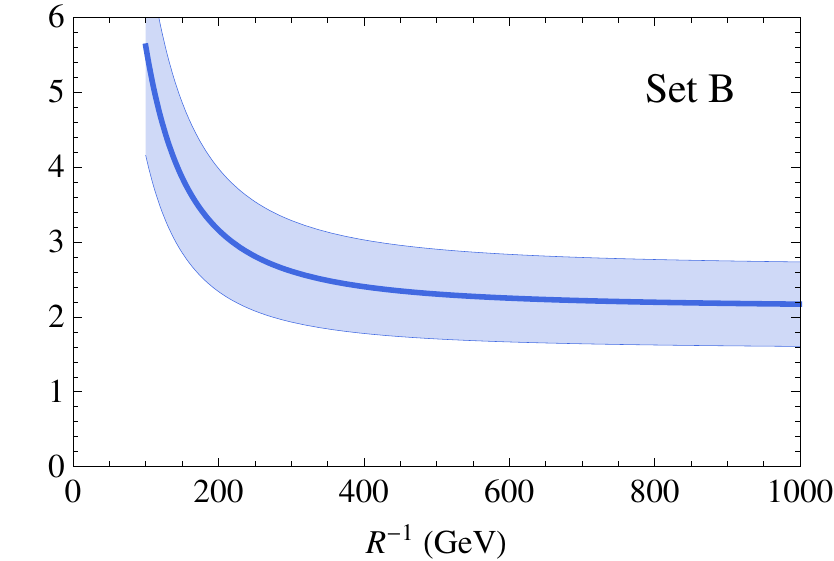}
\hspace*{0.4cm} \includegraphics[width=0.233\textwidth] {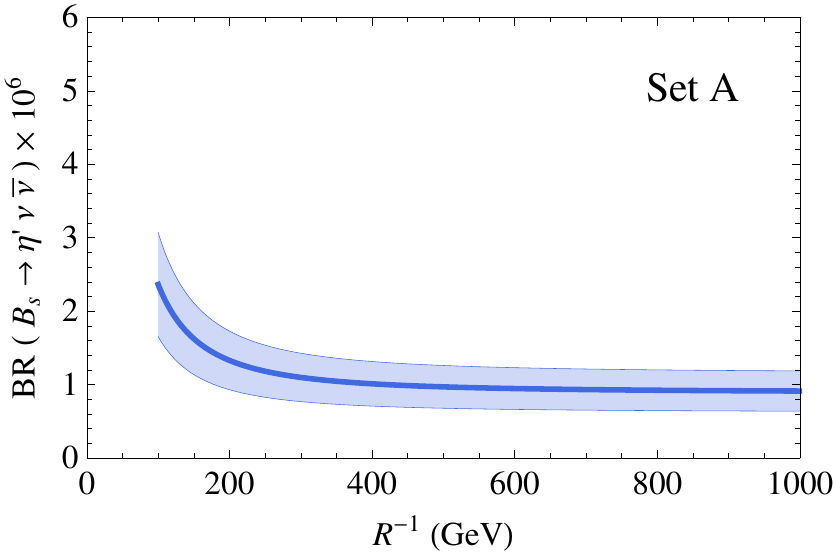}\hspace*{-0.2cm}
 \includegraphics[width=0.225\textwidth] {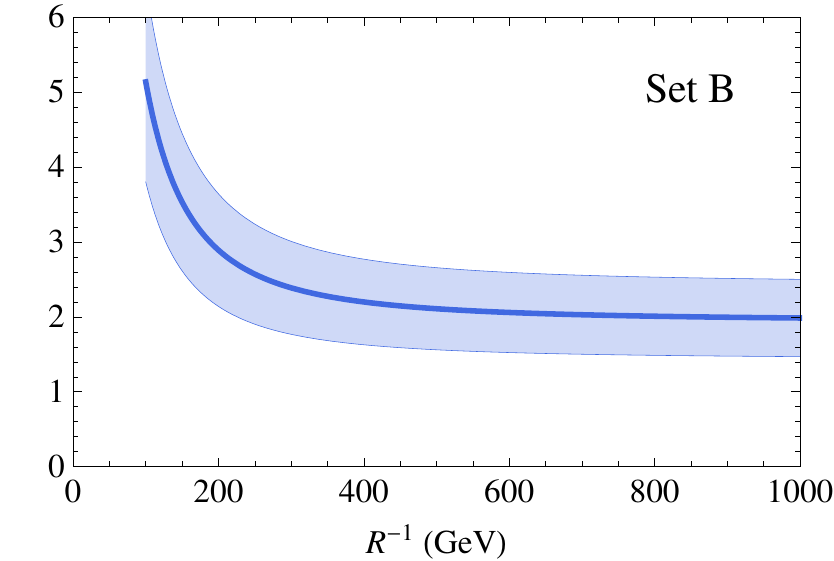}
\caption{
Differential decay rates ${dBR(B_s \to \eta^{(\prime)} \nu \bar \nu) / dx}$ as a function of ${x={E_{miss} / M_{B_s}}}$ (upper plots),
and branching fractions  $BR(B_s \to \eta^{(\prime)} \nu \bar \nu)$ versus $R^{-1}$  (lower plots)
obtained using set A    and B (right) of form
factors. In the upper plots the lowest  region is obtained in  SM;
 the intermediate (blue) one for $1/R=500$ GeV, the uppermost (green) one for $1/R=200$ GeV. } \vspace*{1.0cm}
 \label{etann}
\end{figure*}

The differential branching fraction   is
plotted in Fig. \ref{etann} in the case of $\eta$ and $\eta^\prime$ ,
respectively, for the  sets A and B of form factors. In the Standard Model one predicts:
\be
{ BR}(B_s \to \eta \nu \bar \nu) =
\left\{
\begin{array}{lll}
 (0.95 \pm 0.2) \times 10^{-6} \hskip 0.3cm {\rm set \, A} \\
 (2.2 \pm 0.7)  \times 10^{-6} \hskip 0.5cm {\rm set\, B} \\
 (2.9 \pm 1.5)  \times 10^{-6} \hskip 0.5cm {\rm set \, C} \hskip 2 pt \label{br-eta-nn}
\end{array}
\right.
\ee
\be
{BR}(B_s \to \eta^\prime \nu \bar \nu) =
\left\{
\begin{array}{lll}
 (0.9 \pm 0.2) \times 10^{-6} \hskip 0.4cm {\rm set \, A} \\
 (1.9 \pm 0.5)  \times 10^{-6} \hskip 0.4cm {\rm set\, B} \\
 (2.4 \pm 1.3)  \times 10^{-6} \hskip 0.4cm {\rm set \, C} \hskip 2 pt \label{br-etap-nn}
\end{array}
\right.
\ee
These predictions can be compared to  the present upper bounds for the corresponding $B\to K \nu \bar \nu$ modes, quoted by HFAG:
${ BR}(B^0\to K^0 \nu \bar \nu)<160 \times 10^{-6}$ and
${BR}(B^+\to K^+ \nu \bar \nu)<14 \times 10^{-6}$  \cite{HFAG}, which are  compatible with the SM expectations \cite{scrimieri1}.

In the ACD model,  the coefficient is slightly enhanced:
for example $\displaystyle{C_L^{ACD} \over C_L^{SM}} \simeq 1.10$  at $1/R=300$ GeV.
However, the $1/R$ dependence of the branching fractions  is obscured by the hadronic
uncertainty, as shown in Fig. \ref{etann}.  The distribution in missing energy
shows an enhancement at small values of $x$ which becomes  sizeable at  the lowest  values of the $1/R$ range, $1/R \sim 250$ GeV.
The presence of  two neutrinos in the final state makes  the measurements of these modes a challenge that can be faced at
 an $e^+ e^-$  superB factory operating at the $\Upsilon(5S)$ peak.

\section{The $bs$ UT triangle in ACD}

To complete our discussion about   $B_s$ physics  in the ACD model, we  turn to the {\it bs} unitarity triangle (UT)  and,  in
particular, to the weak phase $\beta_s$ defined as:
$\beta_s=Arg\left[-\displaystyle{V_{ts}V_{tb}^* \over V_{cs}V_{cb}^*}\right]$.
Being ACD a Minimal Flavour Violation  model,  the CKM matrix has the same structure as in the SM: it is constrained to be unitary and is described in
terms of four parameters, one of which is a complex phase:
 $\rho$, $\eta$, A and $\lambda$ in the
Wolfenstein parameterization \cite{Wolfenstein:1983yz}.   At {\cal
O}($\lambda^3$) only $V_{ub}$ and $V_{td}$ have a complex phase,  while at {\cal
O}($\lambda^4$)  also $V_{ts}$ is complex and its argument
identifies  $\beta_s$ \cite{Buras:1994ec}.  

As  it is well
 known,  the various unitarity conditions are represented as
 triangles in the  complex  plane.
 Even though the CKM matrix has the same structure in  ACD and  SM,  and, in particular, its phase is in both cases the
 only source of CP violation, it may happen that the various CKM
 elements differ in the two models and therefore the UT triangles
 do not coincide. 
 Deviations could be expected  for  elements  extracted from
 loop-induced processes, where the tower of KK modes may play a
 role;  for elements  obtained
 from tree level decays no difference is expected, as for  $|V_{us}|$, $|V_{ub}|$ and $|V_{cb}|$.
For the {\it db} triangle defined by the relation: 
$V_{ud}V_{ub}^*  + V_{cd}V_{cb}^*  + V_{td}V_{tb}^*  = 0$,
only the side $R_t=\displaystyle{|V_{td}V_{tb}^* | \over
|V_{cd}V_{cb}^*|}$ depends on $R^{-1}$ through the dependence of
$|V_{td}|$ coming from the KK contribution to the $B_d^0
-{\bar B}_d^0$ mixing  \cite{buras}. Using such a
result, together with the measured value of $\displaystyle{ |V_{ub}|
\over |V_{cb}|}$,  the variation of 
$\bar \rho=\left(1-{\lambda^2 \over 2} \right) \rho$,
 $\bar \eta=\left(1-{\lambda^2 \over 2} \right) \eta$,
and of  $\gamma$ (the phase of $V^*_{ub}$) has been obtained  \cite{buras}, finding  that
the {\it db} triangle in the ACD model could be slightly different than in the SM. 
\begin{figure*}[ht]
\includegraphics[width=0.28\textwidth] {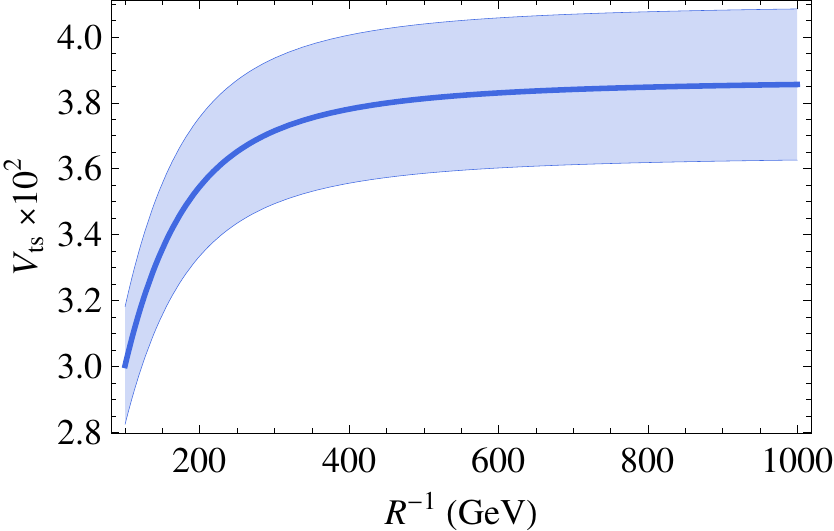} \hspace*{1cm}
\includegraphics[width=0.28\textwidth] {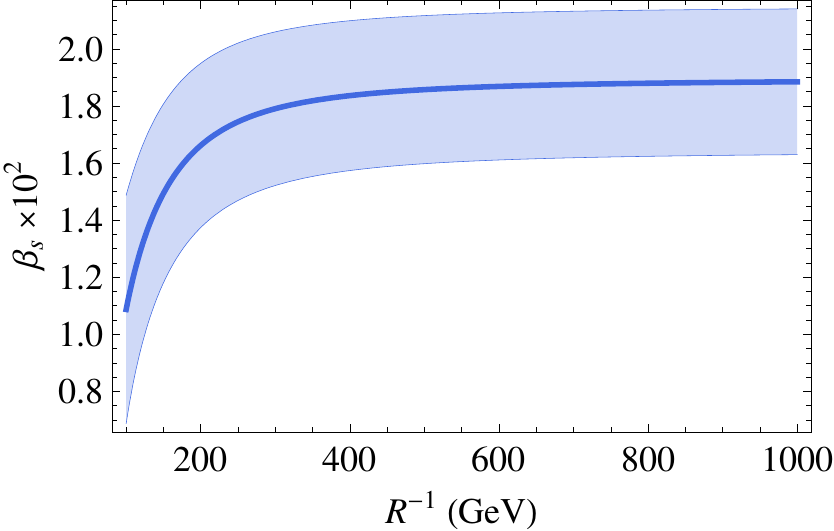}
\caption{ CKM  element $V_{ts}$ (left)  and phase $\beta_s$ (rad)  (right) versus $R^{-1}$.} 
 \label{vts}
\end{figure*}

The {\it bs}  triangle stems from the unitarity condition involving
CKM elements relevant for $B_s$ decays,  and is defined by the  relation:
 \be V_{us}V_{ub}^*  + V_{cs}V_{cb}^* + V_{ts}V_{tb}^*  = 0 \label{bsUT}\,\,. \ee
Analogously  to $V_{td}$, $V_{ts}$ inherits the
$R^{-1}$ dependence from $B_s^0 -{\bar B}_s^0$ mixing \cite{buras}: 
$\Delta M_s \propto S(x_t,1/R)|V_{ts}|^2$, where the
proportionality factor does not depend on $R^{-1}$ and is common to ACD and SM. The function $S(x_t,1/R)$ takes
into account  the SM contribution, $S_0(x_t)$,  and
the contribution of  the KK modes:
$S(x_t,1/R)=S_0(x_t)+\sum_n S_n(x_t,1/R)$, with:
\begin{widetext}
\bea \sum_n S_n(x_t,x_n)= &-&{x_t(5-10x_t+x_t^2) \over 8
(x_t-1)^2}+ {\pi M_W R \over 2} \Big\{-{x_t(3x_t-1) \over 2(x_t^2-1)^3 }
J(R,-1/2)+{(1-3x_t+7x_t^2-x_t^3) \over 2 (x_t-1)^3}J(R,1/2)\nonumber \\ 
&-& {(-3+6x_t-x_t^3) \over 2 x_t(x_t-1)^3}J(R,3/2)
+{1 \over 2 x_t} \int_0^1 dy \,[x_t+3(1+x_t)y]\sqrt{y} \,
{\rm ctanh}(\pi M_W R \sqrt{y}) \Big\}
 \eea 
 \end{widetext}
and
\be J(R,\alpha)=\int_0^1 dy y^\alpha \big[{\rm ctanh}(\pi M_W R \sqrt{y}) -x_t^{1+\alpha} {\rm ctanh}(\pi m_t R \sqrt{y})\big] . 
 \ee
As a consequence one finds: 
\be |V_{ts}|_{ACD}=|V_{ts}|_{SM}
\sqrt{\displaystyle{S_0(x_t) \over S(x_t,1/R)}} \,\,\, .\ee 
In Fig.\ref{vts} we show the dependence of $|V_{ts}|_{ACD}$ on $R^{-1}$
using the quoted value of $|V_{ts}|_{SM}$. 
Since  at {\cal
O}($\lambda^4$) $V_{ts}=-A \lambda^2+{A \lambda^4 \over 2} [1-2
\rho-2i \eta]$, it turns out that
$ \beta_s= \arctan \left[ \displaystyle{2 \lambda^2 \eta \over
2-\lambda^2(1-2 \rho)} \right]$. Using the dependence of $\rho$ and $\eta$ (or  $\bar \rho$ and  $\bar \eta$),
also the $\beta_s$ dependence on $R^{-1}$  can be  obtained;  it  is
depicted in Fig. \ref{vts}. The SM result  for $\beta_s$ is small: $\beta_s
\simeq 0.019$ rad,  and the dependence on the compactification scale $R^{-1}$  further reduces this
value. As  mentioned in the Introduction,  preliminary Tevatron
results   point towards larger values of $\beta_s$, so that
our analysis  supports the conclusion that a confirmation of the measurement  of a large phase in the $B_s$ mixing 
would point towards new Physics models different from MFV scenarios.

\begin{figure}[ht]
\vspace*{-15.5cm}
\hspace*{-1cm}
\includegraphics[width=0.8\textwidth] {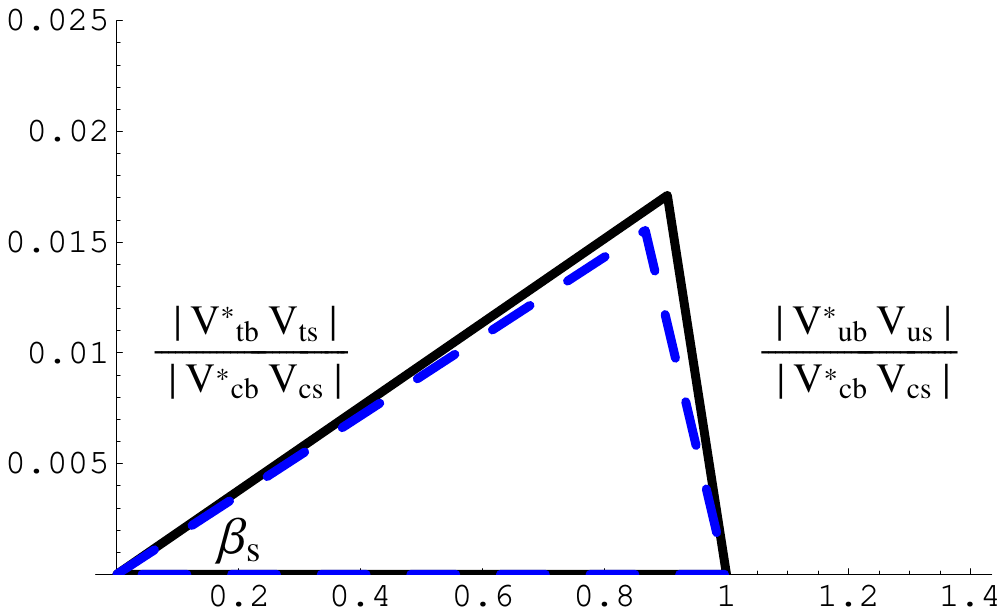}
\caption{ Rescaled ${\it bs}$ UT triangle in the SM (continuous line) and the ACD model   at   $R^{-1}=300$ GeV (dashed line).} 
\label{ut}
\end{figure}

In Fig. \ref{ut}  the rescaled ${\it bs}$ UT triangle in the ACD model is displayed  at the representative compactification scale   $R^{-1}=300$ GeV.

\section{Conclusions}

Together with  other rare $B_s$ decays,  the loop-induced transitions $B_s \to \eta^{(\prime)} \ell^+ \ell^-$ and
$B_s \to \eta^{(\prime)} \nu \bar \nu$  must be included in the Physics programmes of experiments  aimed at performing  tests of  the Standard Model and  at
searching  signals of new Physics   like those  related to  extra dimensions. Within
SM, using the flavour scheme for the $\eta-\eta^\prime$ mixing and the angle  determined by the KLOE Collaboration from radiative $\phi$
decays, we  have predicted decay widths and distributions of these rare $B_s$ channels.
The results for  the branching fractions of modes with two charged leptons are
${\cal O}(10^{-7})$ for $\ell=e,\mu$ and  ${\cal O}(10^{-8})$ for $\ell=\tau$, suggesting that they  are within the reach of  facilities like a SuperB factory. In the case of   the final states with two neutrinos,  the branching ratios are of
${\cal O}(10^{-6})$.  The uncertainty in the numerical results is
dominated by the error of  the hadronic form factors: in order to be sensitive to  compactification scales   $1/R \sim 350-400$ GeV in the ACD model  the error in the form factors should be reduced by about   a factor of two.

A  remark concerns the issue of the  $\eta-\eta^\prime$ mixing.  In addition to modes  with light mesons,  processes involving heavy mesons
  provide information on this mixing. 
Examples  are  the $J/\psi$ radiative decays:  $J/\psi \to \eta^\prime \gamma$ and $J/\psi \to \eta \gamma$, which are
sensitive to the glue content of $\eta$ and $\eta^\prime$ \cite{Ball:1995zv}.  
Although experimentally challenging and considered mainly for different purposes, also the ratio $\displaystyle{ {BR}
(B_s \to \eta^\prime \ell^+  \ell^-) \over {BR} (B_s \to \eta
\ell^+   \ell^-)}$  represents a possibility to determine $\varphi$ and to  look at the gluonic content of $\eta^\prime$. 
 A discrepancy in the comparison of the determination of the same quantity from other
channels, e.g.  from the ratio $\displaystyle{ { BR} (\phi \to
\eta^\prime \gamma) \over {BR} (\phi \to \eta \gamma)}$, would
signal   peculiar effects 
in  $B_s \to \eta^{(\prime)}$ modes.

 \vspace*{1.5cm} \noindent {\bf Acknowledgments} \\
 
\noindent
 FDF thanks Thorsten Feldmann and Tobias Hurth for collaboration on light-cone QCD sum rules
 in the SCET framework.
 This work  was supported in part by the EU
contract No. MRTN-CT-2006-035482, "FLAVIAnet".

 
\end{document}